\documentstyle[11pt,amssymb,amsfonts,amsthm,amssymb,amscd,amstext,epsfig]{article}

\def\ben{\begin{equation}}
\def\een{\end{equation}}

 \let\b=\beta   
  \let\q=\theta 
\let\l=\lambda     
 \let\t=\tau

\let\la=\label

\let\pa=\partial
\def\be{\begin{equation}}
\def\ee{\end{equation}}
\def\ba{\begin{array}}
\def\ea{\end{array}}

\def\dalemb#1#2{{\vbox{\hrule height .#2pt
        \hbox{\vrule width.#2pt height#1pt \kern#1pt
                \vrule width.#2pt}
        \hrule height.#2pt}}}

\newcommand{\bea}{\begin{eqnarray}}
\newcommand{\eea}{\end{eqnarray}}
\newcommand{\tr}{{\rm tr} }
\newcommand{\Tr}{{\rm Tr} }

\def\R{{{\Bbb R}}}

\def\Im{{{\frak{Im}}}}
\def\Re{{{\frak{Re}}}}

\def\6{\partial}

\def\b{\beta}
\def\l{\lambda}
\def\tr{{\rm tr}}

\def\ra{\rangle}
\def\la{\langle}

\begin{document}
\begin{flushright}
NSF-KITP-06-71\\
hep-th/0610103
\end{flushright}

\begin{center}
\vspace{1cm} { \LARGE {\bf Thermal ${\mathcal N} = 4$ SYM theory as a 2D Coulomb gas}}

\vspace{1.1cm}

Sean A. Hartnoll${}^1$ and S. Prem Kumar${}^2$

\vspace{0.8cm}

{\it ${}^1$ KITP, University of California Santa Barbara \\
 CA 93106, USA } \\
 {\tt hartnoll@kitp.ucsb.edu} \\

\vspace{0.3cm}

{\it ${}^2$ Department of Physics, University of Wales Swansea \\
        Swansea, SA2 8PP, UK } \\
{\tt s.p.kumar@swansea.ac.uk }

\vspace{2cm}

\end{center}

\begin{abstract}

We consider ${\mathcal N} = 4$ supersymmetric Yang-Mills theory
with $SU(N)$ gauge group at large $N$ and at finite temperature on
a spatial $S^3$. We show that, at finite weak 't Hooft coupling,
the theory is naturally described as a two dimensional Coulomb gas
of complex eigenvalues of the Polyakov-Maldacena loop, valued on
the cylinder. In the low temperature confined phase the
eigenvalues condense onto a strip encircling the cylinder, while
the high temperature deconfined phase is characterised by an
ellipsoidal droplet of eigenvalues.

\end{abstract}

\pagebreak
\setcounter{page}{1}

\section{Introduction}

\subsection{Known results on the phase structure of thermal ${\mathcal
    N} = 4$ SYM}

At present ${\mathcal N} = 4$ supersymmetric Yang-Mills (SYM)
theory provides the canonical example of an explicit gauge
theory/string theory duality at large $N$ \cite{Maldacena:1997re}.
The extension of the correspondence to finite temperature
\cite{Witten:1998qj} suggests that the duality may allow the phase
structure of the theory to be fully mapped out as a function of
temperature and coupling. From a gravitational perspective, the
finite temperature theory furnishes a unique window into black
hole physics. A key question, for instance, is whether stringy
corrections fundamentally alter the nature of black holes in
strongly curved spacetimes, or whether weakly coupled plasmas are
qualitatively similar to black holes.

The determination of the phase structure of ${\mathcal N} = 4$ SYM
theory at finite temperature on a spatial $S^3$ began with
Witten's identification \cite{Witten:1998qj, Witten:1998zw} of the
first order Hawking-Page transition in the bulk as a strong
coupling deconfinement transition. It was later noticed that a
similar first order transition
\cite{Sundborg:1999ue, Aharony:2003sx} occurs at a critical temperature
in the free field theory on $S^3$ at large $N$. In both of these
transitions, at zero and strong coupling, the free energy of the
system jumps from ${\mathcal O}(1)$ to ${\mathcal O}(N^2)$ at the
critical temperature, as the theory deconfines. This suggests that
the phase transitions should be identified via a phase boundary
running from zero to strong coupling. In fact this boundary turned
out to have some structure of its own, as is described in
\cite{Aharony:2003sx}.

The deconfinement transition at weak coupling was given an elegant
treatment in \cite{Aharony:2003sx}, where the theory was
systematically shown to reduce to a unitary matrix model for the
Polyakov loop. This is the holonomy of the $SU(N)$ gauge field
around the Euclidean time circle of the thermal field theory
\be
U = P e^{i \oint A_0 d\t} \,.
\ee
In the low temperature confining phase $\la \Tr U^k \ra = 0$,
while at high temperatures these traces are nonvanishing. Such
behaviour of these traces implies that, in the large $N$ limit,
the distribution of eigenvalues of the Polyakov loop is uniform in
the confining phase and nonuniform at high temperatures.

\subsection{The role of scalar fields}

This work was motivated by two open questions from previous
treatments:
\begin{itemize}
\item The order parameter for the weak coupling
deconfinement transition is the eigenvalue distribution of the
Polyakov loop. However, the order parameter which can be computed
at strong coupling, and which is natural from the viewpoint of the
dual string theory, is the Polyakov-Maldacena loop
\cite{Maldacena:1998im}. To compare the phases at strong and weak
coupling, we would like to use the same order parameter.

\item At zero 't Hooft coupling all fields except the zero mode of
$A_0$ (or equivalently, the Polyakov loop), are massive and can be
integrated out. This is possible because all the scalars have
positive conformal mass term. However, we shall see that quantum
effects at nonzero 't Hooft coupling $\lambda$ cause condensation
of the scalar fields as well as $A_0$ \cite{Hollowood:2006xb}.
\end{itemize}

These two questions are related. Recall that the
Polyakov-Maldacena loop \cite{Maldacena:1998im} is
\be
W = P e^{i \oint [A_0 + i \Phi_J \theta^J(\t)] d\t} \,,
\ee
where $\theta^J$ is a unit vector in $\R^6$. We can see that if
$A_0$ and $\Phi_J$ have commuting nonvanishing vacuum expectation
values, then the eigenvalue distribution of $i A_0 +
\Phi_J$ is well defined and will be directly related to the
computation of $\la \tr W^k \ra$.

In this letter we study the eigenvalue distribution of $\Phi_J + i
A_0$ at finite weak coupling as a function of temperature. This is
a distribution on the complex cylinder $\R \times S^1$. The
resulting two dimensional framework is the necessary
generalisation to nonzero coupling of studies of the free theory,
such as \cite{Aharony:2003sx}, which were phrased in terms of the
eigenvalue distribution of $A_0$ on $S^1$.

In the limits of low and high temperature the eigenvalue
distribution is given precisely by a 2D Coulomb gas in an external
potential. At intermediate temperatures, the interaction potential
has deviations from the Coulomb law. We compute the distributions
analytically in these limits: the results are constant density
droplets that are shown in figures 1 and 2 below.

\section{The two dimensional Coulomb gas}

\subsection{One loop effective potential}

To understand the phase structure of the theory at finite, weak 't
Hooft coupling we need the quantum effective potential generated
at finite temperature. Such an effective potential can be a
function of all homogeneous modes in the theory on $S^3\times
S^1$. In particular it will depend on the zero modes of the six
scalar fields of the ${\cal N}=4$ theory and the Polyakov loop (or
$A_0$). Fortunately, this potential has very recently been
computed at one loop order
\cite{Hollowood:2006xb}. Several consistency checks were
also performed in that paper. There are three contributions: the
classical term and then the bosonic and fermionic fluctuation
determinants obtained by integrating out the respective
Kaluza-Klein harmonics on $S^3\times S^1$. The potential turns out
to be a gauge invariant function depending only on the eigenvalues
$\{\phi_{Jp}\}$ of the adjoint scalars and those of the time
component of the gauge field $\{\theta_p\}$
\be
S_{\rm eff}[\varphi_{p}, \theta_p]\;=\; S^{(0)}+S^{(1)}_{\rm b} +
  S^{(1)}_{\rm f} \,. \label{full}
\ee
 The classical term is
\be
S^{(0)}=\beta R\pi^2{N \over \lambda}\; \sum_{p=0}^{N-1}
\varphi_{p}^2\,,
\ee
which is just the scalar mass term due to conformal coupling
to the curvature of the $S^3$ of radius $R$. Furthermore,
$\lambda=g^2_{YM}N \ll 1$ is the 't Hooft coupling and $T={1\over
\beta}$, the temperature. The reason it is possible to have
simultaneously diagonal scalar and $A_0$ background values can be
traced to the ${\cal N}=4$ classical potential and the fact that
it allows a Coulomb branch moduli space on ${\mathbb R}^{3,1}$. In
the presence of diagonal background fields, the off-diagonal
fluctuations are massive and appear quadratically in the action.
This allows them to be consistently integrated out with vanishing
expectation values.

Integrating out the gauge fields, scalar and ghost fluctuations
yields the bosonic contribution to the one loop potential
\bea
\lefteqn{S_{\rm b}^{(1)}\;= \;\sum_{p,q=0}^{N-1}\Big(
\beta\,C_{\rm b}(\varphi_{pq})-\frac{1}{2}\log \left[\cosh \b
  \sqrt{\varphi_{pq}^2} -
  \cos \q_{pq}
\right] -\frac{1}{2}\log 2}
\nonumber \\
& &+\sum_{\ell=0}^{\infty}(2 \ell+3)(2\ell+1)\log
\left[1-e^{-\beta
\sqrt{ (\ell+1)^2\,R^{-2}\,+\varphi_{pq}^2}+i\theta_{pq}}\right]+{\rm{c.c}}
\Big)
\,\,, \label{sbosonic}
\eea
and
\bea
S_\text{f}^{(1)} = \sum_{p,q=0}^{N-1}
\Big(\beta\,C_{\rm f}(\varphi_{pq})-\sum_{\ell=1}^\infty
4\ell(\ell+1)\log
\left[1+e^{-\beta
\sqrt{ (\ell+{\textstyle \frac{1}{2}})^2\,R^{-2}\,+\varphi_{pq}^2}+i\theta_{pq}}\right]+{\rm{c.c}}
\Big)
\,.
\label{sfermionic}
\eea
In these expressions $\theta_{pq} = \theta_p - \theta_q$ and
\be
\varphi_p^2 = \sum_{J=1}^6 \phi_{Jp}^2 \,, \qquad \varphi_{pq}^2
= \sum_{J=1}^6 (\phi_{Jp} - \phi_{Jq})^2 \,.
\ee
The functions $C_{\rm b}$ and $C_{\rm f}$ are the bosonic and
fermionic Casimir energy contributions in the presence of
background expectation values for the adjoint scalar fields
\cite{Hollowood:2006xb}. Their sum yields the zero
temperature one loop effective potential of ${\cal N}=4$ theory on
$S^3$. For present purposes it will suffice to know that this
potential has the expansion
\be
C_{\rm b}(\varphi_{pq})+C_{\rm f}(\varphi_{pq}) = {1\over
R}\left[\frac{3}{16} +(R\varphi_{pq})^2 (\log 2-\frac{1}{4})+{\cal
O}(R^4\varphi_{pq}^4)+\cdots
\right]
\ee
near the origin.

The problem now is to find the eigenvalue distribution that
minimises the effective potential (\ref{full}).

\subsection{Instability of the ring configuration and short distance repulsion}

Previous works have considered condensation of the gauge field
eigenvalues $\theta_p$ while assuming that the scalar eigenvalues
remained zero: $\phi_{Jp} = 0$. We will instead look for general
distributions in which both sets of eigenvalues can condense. Let
us introduce the complex field
\be\label{eq:defz}
z_p = \frac{\b \phi_p + i \theta_p}{2} \,.
\ee
The $z_p$ live on a cylinder: $\Re\;z_p$ ranges from $-\infty$ to
$+ \infty$ but $\Im\;z_p$ is periodic with range $\pi$.

In our definition of the complex variables $z_p$ and in what follows
we will omit $SO(6)$ vector indices on $\phi_{Jp}$ and think of it as the
radial mode in $SO(6)$ space. A non-zero radial mode implies symmetry
breaking. This is possible on $S^3 \times S^1$ due to the large $N$
limit which plays the role of the thermodynamic limit
\cite{Hollowood:2006xb}. Hence we set $\varphi_p=\phi_p$ and
$\varphi_{pq}=\phi_p-\phi_q$.

The first statement we can make is that although $\phi_p = 0$ is
always a solution, call it the ring configuration, it is never a
stable configuration at finite 't Hooft coupling. This is due to a
strong repulsive force between the eigenvalues at short distances,
originating from the first line of (\ref{sbosonic}). The term may
be rewritten as
\be\label{eq:exposum}
\frac{1}{2} \left( \log 2 + \log \left[\cosh
\b \sqrt{\varphi_{pq}^2} - \cos \q_{pq} \right] \right)
 = \log 2 + \log | \sinh z_{pq} |\,.
\ee
This term reduces to the familiar $\log \sin
\frac{\q_{pq}}{2}$ when $\phi_p=0$. As $z_p \to z_q$ we see that
the eigenvalues experience a 2D Coulombic repulsion with potential
$- \log | z_{p} - z_{q} |$. At large $N$ there is no other term in
the action (\ref{full}) that can compete with this term at short
distances and therefore the eigenvalues always spread out into two
dimensions. Let us see this explicitly.

\subsection{Low temperature limit}

Consider first the low temperature limit of the confining phase.
That is, take $\frac{R}{\b} \ll 1$. The action (\ref{full}) at low
temperatures becomes
\be\label{eq:normal}
S_{\rm R T \ll 1} = N^2 \left[ \frac{3\b}{16 R} - \log 2 \right] +
\left[ \frac{N \pi^2 R}{\b \l}
\sum_{p=0}^{N-1} (z_p + \bar z_p )^2 - \sum_{p,q=0}^{N-1} \log
| \sinh z_{pq} |\right] \,.
\ee
As well as the terms in the action which are exponentially
suppressed at low temperatures, we have also dropped all the
Casimir energy terms from (\ref{sbosonic}) and (\ref{sfermionic}).
This will be valid if $R \phi_{Jp} \ll 1$. We will see a
posteriori that this is true for the solution we are about to
present.

The action in (\ref{eq:normal}) is an elegant reformulation of the
problem as a Coulomb gas in two dimensions with an external
potential. This system is closely related to normal matrix models.
The interaction potential $\log |
\sinh z_{pq} |$ is precisely the Coulomb potential on a cylinder.

In writing down the equations of motion following from
(\ref{eq:normal}) it is convenient to pass to the continuum large
$N$ limit. This is done by introducing an eigenvalue distribution
\be\label{eq:continuum}
\sum_{q = 0}^{N-1} \to N \int d^2z \rho(z, \bar z) \,.
\ee
The continuum limit implies the normalisation
condition\footnote{We work with the convention that $d^2z = 2 dx
dy$, where $z = x + iy$.}
\be\label{eq:normalise}
\int d^2z \rho(z, \bar z) = 1 \,,
\ee
while the equations of motion become
\be\label{eq:eom}
z + \bar z = \frac{\b \l}{2 \pi^2 R} \int d^2z'
\rho(z',\bar z')
\coth(z-z') \,.
\ee

As is standard for these types of system, we can immediately
derive from (\ref{eq:eom}) that the eigenvalue distribution is
everywhere either a fixed constant, or it is zero. This follows
from acting on (\ref{eq:eom}) with $\bar \pa$, and gives
\be
\rho(z, \bar z) = \left\{
\begin{array}{c}
\frac{\pi R}{\beta \lambda} \quad \text{if} \quad z \in B \\
0 \quad \text{if} \quad z \notin B
\end{array} \right. \,.
\ee
Here $B$ is a subset of the complex cylinder. In other words, the
eigenvalue distribution takes the form of a uniform density
`droplet' on the cylinder. The equations of motion (\ref{eq:eom})
therefore imply that we need to find a domain $B$ such that
\be\label{eq:eom2}
z + \bar z = \frac{1}{2\pi} \int_B d^2z' \coth (z-z') \,.
\ee

From symmetry considerations it is not difficult to see what form
the droplet has to take. The external force acting on the
eigenvalues acts parallel to the real axis and pushes the
eigenvalues towards the imaginary axis. The repulsive force
between the eigenvalues must balance this effect to lead to a
finite width band. This is illustrated in figure 1 below. We will
show momentarily that a band of constant width $2A$ does indeed
solve (\ref{eq:eom2}), for all $A$. Given this fact, the
normalisation condition (\ref{eq:normalise}) determines the width
$A$ of the band to be
\be
A = \frac{\beta \lambda}{4 \pi^2 R} \,.
\ee

\begin{figure}[h]
\begin{center}
\epsfig{file=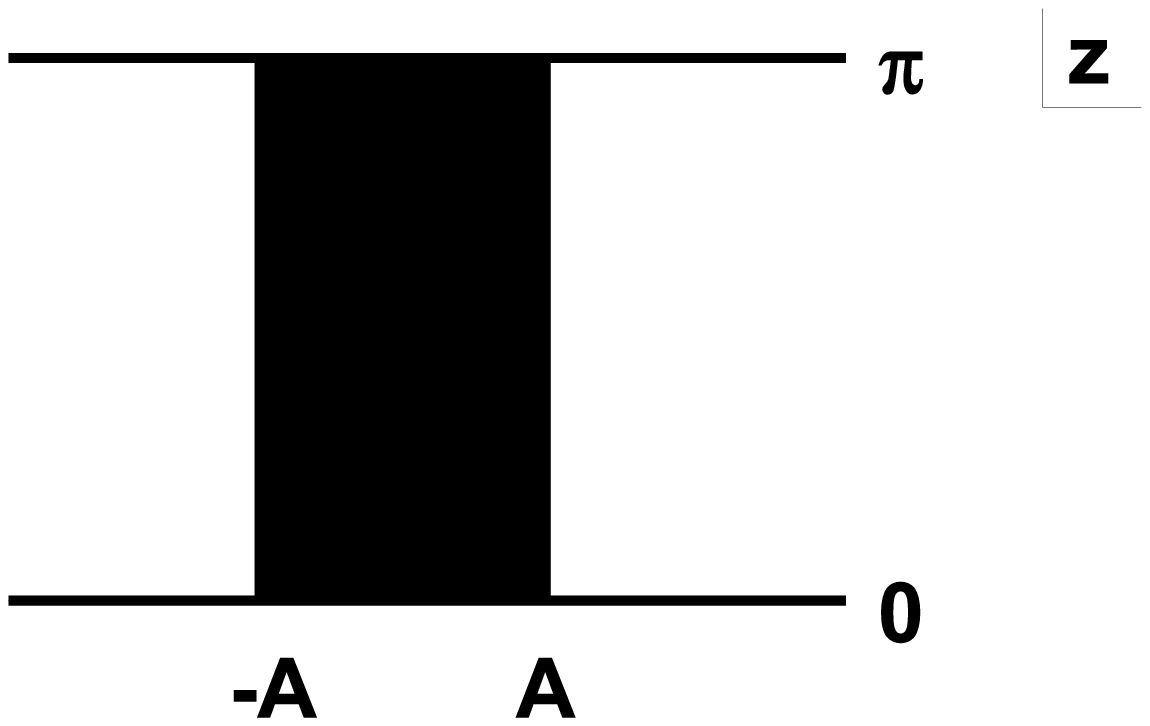,width=8cm}
\end{center}
\noindent {\bf Figure 1:} The low temperature distribution of
eigenvalues in a band of width
$2A = \frac{\beta \lambda}{2 \pi^2 R}$ on the $z$ cylinder.
\end{figure}

The equation of motion (\ref{eq:eom2}) is checked by doing the
double integral. Set $z' = x + i y$ so that the integral becomes
\bea
\frac{1}{2\pi} \int_B d^2z' \coth (z-z') & = &
 \frac{1}{\pi} \int_{-A}^{A} dx \int_0^{\pi} dy \coth (z - x - i
 y) \nonumber \\
  & = & \int_{-A}^{(z+\bar z)/2} dx - \int_{(z+\bar z)/2}^{A} dx
  \nonumber \\
  & = & z + \bar z \,.
\eea
In going from the first to the second line here, one needs to be
careful to keep track of the branch cut in $\log \sinh$.

Given the band solution, we can justify the low temperature
approximations made to obtain the action (\ref{eq:normal}). We see
that $R \phi \sim \lambda \ll 1$. It follows that evaluated on the
solution, the Casimir energy terms are suppressed by a power of
the 't Hooft coupling relative to the terms kept in
(\ref{eq:normal}). It furthermore follows that because the spread
$R \Delta \phi \sim
\l$ makes no reference to the temperature $\beta$, the condensate
has a finite zero temperature limit. Consistent with confinement, and
with strong coupling results,
the band solution yields $\la\Tr W^k \ra=\la\Tr U^k\ra=0$ due to 
cylindrical symmetry of the distribution which causes all phases to
average out to zero.

Having established that the finite band solves the equations of
motion, we can check that indeed the configuration has lower
action than the ring $z+\bar z = 0$, which we argued was unstable.
Evaluating the action (\ref{eq:normal}) on the solution in the
continuum limit gives
\be
S_{\text{band}} = \frac{N^2 \b}{R} \left[\frac{3}{16} -
\frac{\lambda}{12 \pi^2} \right] \,.
\ee
The ring solution at low temperatures is uniformly distributed
along the ring, $\q_p = 2 \pi p/N$, which gives the action
\be
S_{\text{ring}} = \frac{N^2 \b}{R} \frac{3}{16} > S_{\text{band}}
\,.
\ee

\subsection{Deconfinement transition and high temperature limit}

If we raise the temperature and consider the effects of the
exponentially suppressed terms in (\ref{sbosonic}) and
(\ref{sfermionic}), then we find that the constant density band
solution we have just described remains a solution. This occurs
because the action may be expanded in terms of $\cos (n
\theta_{pq})$ factors, which give zero when summed over the uniform
distribution on the $\theta$ circle.

The behaviour of these sums of exponentials is identical to that
described in \cite{Aharony:2003sx} for the Polyakov loop matrix
model. Let us recall from that work that when $\b/R$ becomes of
order one, then although the uniform distribution in the $\theta$
direction remains a solution, it is no longer the absolute
minimum. This is the deconfinement temperature. The presence of $R
\varphi_{pq}$ in the exponent, the new ingredient in our computations,
will not change this story qualitatively. It will however play a
role in determining the order of the transition, c.f.
\cite{Aharony:2003sx}.

Above the deconfinement transition, the force between eigenvalues
in the circle direction becomes attractive and the eigenvalue
distribution develops a gap along the $\theta$ circle. We postpone
a careful discussion of the behaviour of eigenvalue droplets near
the transition for future work \cite{us}. The story is complicated
by likely non uniformities in the $\phi$ direction.

A regime that is again analytically tractable is the high
temperature limit, $\frac{R}{\b} \gg 1$. From previous work
\cite{Aharony:2003sx}, our intuition is that at high temperatures
the eigenvalues will clump into a small droplet. Let us take,
simultaneously to the high temperature limit, the limit in which
$\theta_p, \b \phi_q \ll 1$ in the action (\ref{full}). Similarly
to before, we will verify a posteriori that this condition is
satisfied by the solution. It is convenient to again introduce the
complex $z_p$ as we did in (\ref{eq:defz}). The action in these
limits becomes
\be
S_{\rm TR \gg 1} = \frac{N \pi^2 R}{\beta \lambda}
\sum_{p=0}^{N-1} (z_p + \bar z_p)^2 - \sum_{p,q=0}^{N-1} \log |z_{pq}|
+ \frac{\pi^2 R^3}{\b^3} \sum_{p,q=0}^{N-1} \left[-z_{pq}^2 - \bar
z_{pq}^2 + 6 |z_{pq}|^2 \right]
\,.
\ee
Although the first two terms in this expression are na\"\i{}vely
subleading, they become important depending on the 't Hooft
coupling and proximity of the eigenvalues. We now take the
continuum limit as we did at low temperatures
(\ref{eq:continuum}). The same arguments as before imply that the
eigenvalues form a droplet $B$ of constant density
\be
\rho(z,\bar z) = \frac{\pi R}{\b \lambda}
\left[1 + \frac{6 R^2 \lambda}{\b^2}\right]\quad \text{if} \quad z \in B.
\ee
The equation of motion becomes
\be\label{eq:highTeom}
\tanh \mu \, z + \bar z = \frac{1}{2\pi} \int_B d^2z'
\frac{1}{z-z'}
\,,
\ee
where we defined
\be\label{eq:mu}
\tanh \mu = \frac{1 - \frac{2 R^2
\lambda}{\b^2}}{1 + \frac{6 R^2
\lambda}{\b^2}} \,.
\ee
The equation (\ref{eq:highTeom}) has been simplified by the
observation that if the symmetry $z \to -z$ is unbroken, then we
must have $\sum_q z_q =
\sum_q \bar z_q = 0$ on the solution.

Before solving the equation of motion, the normalisation condition
(\ref{eq:normalise}) tells us that the area of the droplet on the
complex plane will be
\be\label{eq:area}
\text{Area(B)} = \frac{1}{12\pi^2} \frac{\b^3}{R^3} \frac{1}{1 + \frac{\b^2}{6 R^2
\lambda}} \,.
\ee
This area goes to zero as $\b/R \to 0$ at high temperatures, and
therefore our initial assumption on $\phi_p$ and $\theta_p$ was
justified.

The droplet shape for the Coulomb gas problem (\ref{eq:highTeom})
is known to be an ellipse \cite{DiFrancesco:1994fh}. This is
illustrated in figure 2 below. The semi-axes of the ellipse are of
length $r e^{-\mu}$ along the real axis and $r e^{\mu}$ along the
imaginary axis, where $\mu$ is given in (\ref{eq:mu}) and $\pi
r^2$ equals the area (\ref{eq:area}). The eccentricity is
therefore controlled by the ratio $\lambda R^2/\beta^2$.
Interestingly, it is precisely this ratio which controls the
validity of perturbation theory at finite temperature on $S^3$.
The one loop effective potential and results we have deduced from
it are valid for all temperatures $T \lesssim {R^{-1}\over
\sqrt{\lambda}}$ \cite{Hollowood:2006xb}.

\begin{figure}[h]
\begin{center}
\epsfig{file=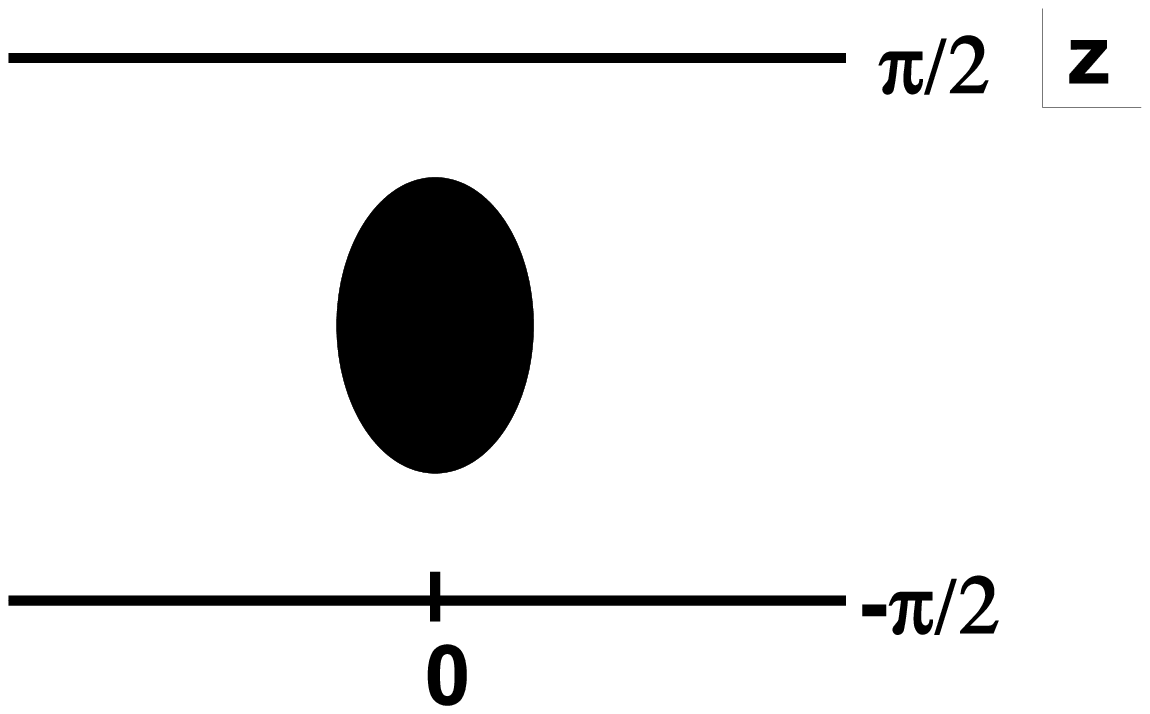,width=8cm}
\end{center}
\noindent {\bf Figure 2:} The high temperature distribution of eigenvalues
in an ellipse (not to scale).
\end{figure}

\section{Discussion}

In this letter we have shown that weakly coupled ${\mathcal N} =
4$ SYM theory at finite temperature on a spatial $S^3$ is
described at large $N$ by an eigenvalue distribution on the
complex cylinder. Remarkably, these eigenvalues are naturally
associated to the Polyakov-Maldacena loop, the order parameter
known to emerge at strong coupling. At high and low temperatures
we determined the distribution by solving an associated 2D Coulomb
gas problem. The deconfinement transition of the theory on $S^3$
corresponds to a `band-to-droplet' transition on the cylinder.   
A key point we emphasised is that the eigenvalues of the scalars as
well as $A_0$ condense. Outstanding questions are the
implications of scalar condensates for the dual string theory and the
relevance of the `band-to-droplet' picture for the emergent Big Black Hole
phase at high temperatures.

It is of immediate interest to understand the droplet picture in the
vicinity of the phase transition, and also whether the strongly coupled
theory may be formulated in the Coulomb gas language.
This could lead to a new approach for counting black hole microstates.
Some progress studying higher traces of the Polyakov-Maldacena loops at
strong coupling was recently made \cite{Hartnoll:2006hr}. The
Coulomb gas framework we have presented has a description in terms of
fermions, suggesting a connection to a different 
droplet picture of certain BPS states recently developed for the
zero temperature theory
\cite{Berenstein:2004kk, Lin:2004nb}.

\section*{Acknowledgements}

We would particularly like to thank Umut G\"ursoy for helpful
comments throughout this work. We have also had useful
conversations with Matt Headrick, Tim Hollowood, Albion Lawrence,
Asad Naqvi and Carlos Nu$\tilde{\rm n}$ez.

This project was begun while SAH was supported by a research
fellowship from Clare College Cambridge. SPK is supported by
a PPARC Advanced Fellowship. This research was supported in part by
the National Science Foundation under Grant No. PHY99-07949.

\appendix

\end{document}